
\tenrm
\magnification=\magstep1

\abovedisplayskip 20pt plus 4pt minus 4pt 
\belowdisplayskip 20pt plus 4pt minus 4pt 

\outer\def\beginhead#1\par{\bigskip\bigskip\message{#1}\centerline{\bf#1}
       \nobreak\bigskip\vskip-\parskip
       \noindent}
\outer\def\beginsection#1\par{\bigskip\bigskip\message{#1}\centerline{\bf#1}
       \nobreak\bigskip\vskip-\parskip\noindent}
\outer\def\beginsubsection#1\par{\message{#1}\leftline{\bf#1}
       \nobreak\medskip\vskip-\parskip\noindent}
\outer\def\beginsubhead#1\par{\message{#1}\leftline{\bf#1}
       \nobreak\vskip-\parskip\noindent}

\baselineskip 15pt
\parskip=0pt plus .2pt minus .2pt   
\widowpenalty=10000   
\clubpenalty=10000    
\raggedbottom   
\predisplaypenalty=300   
\postdisplaypenalty=100  
\doublehyphendemerits=1000
\adjdemerits=1000

\hsize=5.9 true in  
\hoffset=.5 true in 
\vsize=8.3 true in  
\voffset=0.4 true in 

\pageno=1  

\newinsert\pagein
\skip\pagein=0pt
\count\pagein=1000
\dimen\pagein=\maxdimen
\def\page{\begingroup\setbox0=\vbox\bgroup}
\def\endpage{\egroup\insert\pagein{\penalty100
\vbox to\vsize{\unvbox0 }}\endgroup}

\centerline{\bf Vibrational States of the Hydrogen Isotopes on Pd(111)}
\vskip 0.4 true in
\centerline {Steven W. Rick\footnote{*}{Present address: Department of
Chemistry, Columbia University, New York, NY 10027}\footnote{$\dagger$}{contact
author, telephone: (212) 854-5650, fax: (212) 932-1289} and J.D. Doll}
\centerline {\it Department of Chemistry}
\centerline {\it Brown University}
\centerline {\it Providence, RI \ \ 02912}
\vskip 0.4 true in


{\narrower\smallskip\noindent
The ground and excited vibrational states for the three hydrogen
isotopes on the Pd(111) surface have been calculated.  Notable features
of these states are the high degree of anharmonicity, which is most
prominently seen in the weak isotopic dependence of the parallel
vibrational transition, and the narrow bandwidths of these states,
which imply that atomic hydrogen is localized on a particular surface site
on time scales of 100 picoseconds or more.  Experiments to resolve
ambiguities concerning the present system are suggested.

\vskip 0.2 true in
\noindent
\vfill
\eject

\beginsection

Hydrogen atoms chemisorbed on a metal surface[1]
display a variety of phenomena, ranging from localization on a
particular binding site with thermally activated hopping
or incoherent tunneling between the sites
to extended band states, depending on the metal, the face, the
temperature, and the coverage.
A hydrogenic-band model in which the
hydrogen wavefunction is delocalized parallel and localized
perpendicular to the surface was first[2] proposed by Christmann,
{\it et al.}, in 1979 for H on Ni(111)[3].  Puska and co-workers followed
up on this suggestion by calculating the states for hydrogen on a rigid
Ni surface[4]. They find narrow width ground states and excited
states with appreciable width ($\sim$50 meV), particularly for the (110) and
(111) face.  Since then there has
been a number of experiments which reported evidence of band-like
states, mainly along the "troughs" of surfaces such as the (211) face of
body-centered cubic (bcc) metals[5] and the (110)
face of face-centered cubic (fcc) metals[6] or
for the smoothly corrugated (111) face of fcc metals[7,8].
The helium scattering experiments
of Hsu, {\it et al.} find a hydrogen phase with 6-fold
symmetry[8]. The authors explain this phase in terms of hydrogen
being quantum mechanically delocalized over two adjacent 3-fold sites on
the sub-picosecond time scale of the helium scattering event. These
results are for a hydrogen coverage of half saturation
and a temperature range of 140 to 300 K
and the same results are seen for D as for H [9].
The time
scales for hydrogen diffusion have been measured for several metal
surfaces (Ni(100)[10],W(110)[11],Rh(111)[12], and Pt(111)[13]).
At high temperatures,
diffusion shows an Arrhenius dependence indicative of thermally
activated hopping between localized sites, with activation energies
ranging from 0.15 to 0.52 eV.  At temperatures below 100 K, Gomer and
coworkers[11] for H/W(110) and Zhu, {\it et al.}[14], for H/Ni(100) find
a diffusion rate due to tunneling which is independent of temperature.
The diffusion rates at the low temperatures are
very small (10$^{-11}$ to 10$^{-13}$ cm$^2$/sec), indicating very little
wavefunction overlap between adjacent sites and bandwidths of 10$^{-2}$
to 10$^{-11}$ meV [15]. The small bandwidth for H on
Ni(100) is consistent with the band calculations of Puska and co-workers[4]
and high resolution electron-energy loss spectroscopy (HREELS)
experiments[16].

Vibrational spectroscopy such as HREELS provides a direct experimental
probe of the nature of the hydrogen states.  HREELS gives direct
evidence as to whether vibrational excitations correspond to small
vibrations about an equilibrium position as is typically the case
(e.g. H/Ni(100)) or
correspond to transitions to broad bands (H/Cu(110)). An HREELS study
of H/Pd(111) at high coverages (a hydrogen exposure of 2 Langmuirs)
shows two narrow peaks: a parallel vibration at 774 cm$^{-1}$ and a
perpendicular vibration at 998 cm$^{-1}$[17].
In this letter we report the vibrational states for the three
isotopes of hydrogen on Pd(111) in the single adatom limit.  The low
coverage limit presents the best conditions for delocalized states,
since the H-H interaction is expected to be purely repulsive.
Increasing coverage should lead to a higher probability of localized
states.  This supposition is borne out by experiment, in which evidence
of quantum delocalization disappears at high coverages [6,8].
Furthermore, although the widths of the excitations can vary
with coverage, the position of the peak center does not have a strong
coverage dependence.

In the present computations,
the lattice is kept rigid in the minimum energy configuration of the
surface in the absence of any adsorbates.
The magnitude of adsorbate induced reconstructions is examed below.
The clean surface does not reconstruct,
the top surface only contracts by 0.07 ${\rm \AA}$ relative to the bulk
lattice spacing.
The (111) surface is a hexagonal surface with 2 distinct sites: a
tetrahedral site directly above a subsurface atom
and an octahedral
site, over a metal atom in the second subsurface layer.
The two sites are treated as fully degenerate since the potential used
predicts them to be energetically equivalent.
The interaction potential is taken to
be the Embedded Atom Method (EAM) potential of Daw and Baskes [18].
The EAM potential has a form borrowed from
density functional theory
and is determined by fitting to properties of the bulk
system.  Although this potential contains only bulk data, it
successfully reproduces
the phase diagram for H/Pd(111)[19] and predicts the heat of
absorption of hydrogen to be 2.85 eV, which is the experimental
value[20].  This predicted value is the
classical binding energy relative to an isolated atom and is for
hydrogen absorbed on the rigid surface.  Lattice relaxations lower the
energy an additional 0.077 eV. The EAM barrier for surface diffusion is
175 meV.  This barrier is not known experimentally but the barrier to
bulk diffusion is 226 meV[21]. EAM predicts 152 meV for the bulk
barrier, allowing
for relaxation of the palladium atoms.

To find the vibrational states, we use the tight-binding approximation
in which we calculate states localized on a particular binding site and
take the appropriate linear combinations to construct states with the
full 2-dimensional periodicity of the surface (see, e.g., Ref. [22]).
The full periodic Hamiltonian of our system, ${\cal H}$, is
writen as a localized
part, ${\cal H}_0$, plus the corrections, $\Delta$U.
In the localized Hamiltonian, the hydrogen only interacts with the
nearest three surface palladium atoms and the nearest three subsurface
palladium atoms. These six interactions most nearly reproduce the
full potential of the lattice in the region of the 3-fold site.
Three eigenstates of ${\cal H}_0$ are calculated: a ground
state with ${\rm A}_1$ symmetry and excited states with ${\rm A}_1$
and E symmetries.
The states are found by
solving the Schr\"odinger equation on a 3-dimensional mesh
of spacing 0.025 ${\rm \AA}$ and---for the the excited
states---simultaneously projecting out the lower energy states[23].
The accuracy of the ground and E states can be checked using the
Diffusion Monte Carlo
(DMC) method, which solves the Schr\"odinger equation exactly[24].
The E state can be calculated using the DMC method using
the fixed-node approximation, since for the E-symmetry state the fixed
node approximation is exact and the position of the nodal plane is
easily deduced by symmetry.
For the ${\rm A}_1$ state, the fixed-node approximation is not exact.
The Bloch states, $\Psi_{n,{\bf k}}$, with {\bf k} the two-dimensional
wavevector,
are constructed by calculating the overlap between
nearest neighbor surface sites only.
The necessary overlap integrals can be easily done on the same
mesh as is used to solve for the localized states.
The overlap integral between
adjacent sites is less than 0.17, so the tight-binding approximation is
reasonable.

The band centers and widths for the three lowest energy states are shown
on Table I.  The energies are all relative to the classical binding
energy (-2.85 eV).  The states for all three isotopes are very
localized, the only states with a bandwidth larger than 1 meV are the H
excited states.  The width for each state narrows with increasing mass
roughly as e$^{-\sqrt{m}}$[25].

In order to assist with spectroscopic assignments the {\bf k}=0
H bandstates are plotted along the three directions (Fig. 1).
The left side of Fig. 1 shows the states' amplitude as a function of one
surface coordinate, the [1$\bar 2$1] direction, and the [111]
direction perpendicular to the surface, at a constant value for the
other surface coordinate equal to the center of the unit cell.
The right side shows the same wavefunctions as a function of the
two surface coordinates at a constant height of 1${\rm \AA}$ above the surface
plane. The localized nature of the ground state is apparent.  The
excited state A$_1^1$ is much broader and can be identified as
predominantly a perpendicular vibrational excitation, since the nodal surface
is nearly a plane parallel to the surface.
The A$_1^0$ to A$_1^1$ transition will be
dipole active in HREELS, the A$_1^0$ to E transition can be seen only
though impact scattering[26].

Table I also shows the energy differences between the ground and excited
states for the parallel mode (the A$_1^0$ to E transition) and the
perpendicular mode (A$_1^0$ to A$_1^1$) for the three
isotopes. Shown is the energy differences between the centers of the
bands, although for the H isotope the excited states are broad enough
that wide loss peaks might be detectable experimentally.
The ordering of the vibrational peaks is the opposite of Conrad,
{\it et al.}, who assign $\omega_\parallel$=774 cm$^{-1}$ and
$\omega_\perp$=998 cm$^{-1}$
based on
their HREELS experiments at high coverages [17].
The ordering of the parallel and perpendicular modes does vary for
hydrogen on various metals and is a sensitive gauge of
the metal-hydrogen interaction[1].
For examples with three-fold or quasi-three-fold coordination,
H/Ru(0001) [27] and H/Ni(111) [28] have $\omega_\perp > \omega_\parallel$
and H/Pt(111)[29] and H/Rh(110)[30] have $\omega_\parallel > \omega_\perp$,
although the H/Pt(111) assigment is challenged by Feibelman and Hamann [31].
If the experimental assignments for H/Pd(111) are correct,
the EAM potential used here
appears to misrepresent the relative stiffness of the two
vibrations.

The isotope dependence of the computed frequencies is unusual and demonstrates
the importance of anharmonic effects, particularly for the parallel
frequency.  The ratio of $\omega_\parallel$(H)/$\omega_\parallel$(D)
is only 1.19, different from the $\sqrt2$ dependence of
harmonic vibrations.  The ratio of $\omega_\perp$(H)/$\omega_\perp$(D)
is 1.32, closer to $\sqrt2$.
The anharmonic nature of the vibrational
states can be seen by comparing the exact energies with harmonic estimates
(Fig. 2).
The band state energies (the longer horizontal line on the left) is shown
connected (by the dotted line) to the harmonic estimate of the energy of
that state (the smaller horizontal line on the right).  The harmonic
estimate is based on the curvature of the potential minimum of the
absorbed state.
Also visible in Fig. 2 are the bandwidths of the A$_1^1$ and E
states. All the states are lower in energy than the harmonic estimate.
Most notable is the E state of H, which is lower by 90 meV or about
20$\%$.
That the vibrational frequencies are sensitive to features of the
potential other than the curvature of the well bottom should be a
warning to those who might attempt to construct potentials by fitting
harmonic estimates to experimental values of the frequencies.

Most experimental spectra have a isotope dependence closer to $\sqrt2$,
but the weak dependence of the parallel mode has precedence:
HREELS experiments by Mate and Somorjai for H/Rh(111) find only a
very small isotopic shift of about 10 cm$^{-1}$ for a loss feature
believed to be a transition to a delocalized band with E symmetry[7].
HREELS experiments for H and D on Ni(100)[32] and Rh(100)[33] have frequency
ratios of the perpendicular mode
close to $\sqrt2$, although overtones in the H/Rh(100) spectra show noticeable
anharmonicities. For H and D on Pd(100) [34], the $\omega_\perp$ ratio is
about 1.3 and for the Ni(110) and Rh(110) surfaces the ratio are 1.25[35]
and 1.35 [30], respectively.  Calculations
for H/Rh(0001) find isotope ratios for several states all in
the range of 1.41 to 1.46 and harmonic estimates differ from the anharmonic
calculations by, at most, 8$\%$[36].

The H/Pd(111) bandwidths are much narrower and the band gaps larger than those
calculated for H/Ni(111)[4].
The ground state for that system has a width of 4 meV, the first excited
state, which has E symmetry, has a width of 40 meV, and the next excited
state, with A$_1$ symmetry, has a width of 74 meV.
The larger widths of the Ni states is due to
the lower diffusion barrier for H on Ni(111)[18] and
the shorter lattice constant for Ni (3.52${\rm \AA}$ for Ni, 3.89
${\rm \AA}$ for Pd).

{}From the localized tight-binding states,
estimates of both the magnitude of hydrogen
induced reconstructions and time scales for hydrogen tunneling
between adjacent surface sites can be made.
As mentioned previously, lattice distortions lower the classical binding
energy by 77 meV, as predicted by the EAM potential.  This energy is a
factor of ten larger than the largest bandwidth (Table I) and four orders
of magnitude greater than the ground state width.  Thus the small
polaron condition is easily met: the energy gained from self-trapping
the adsorbate by small lattice distortions is greater than the kinetic
energy gained by forming extended bands[37]. Small
polaron theory provides a convenient formalism for describing the
tunneling process from site to site, including the influence of
phonons.
The small polaron Hamiltonian is found by
expanding the potential around the rigid lattice potential,
${\rm V}^{rigid}$({\bf r}), to first order in the phonon coordinates,
$$ {\rm V}^{sp}({\bf r},{\bf Q}) =
 {\rm V}^{rigid}({\bf r})
+ \sum_l V_l^a
 Q_l
\eqno(1)
$$
where
$$
{\rm V}_l^a = { \partial {\rm V}(\bf r, {\bf Q} )
\over \partial {\bf Q}_l } \bigg
\vert_{{\bf r}={\bf r}^a} ,
$$
${\rm V}({\bf r},{\bf Q})$ is the EAM potential energy,
$Q_l$ is the $l^{th}$ perfect lattice (no adsorbate)
phonon coordinate
and {\bf r}$^a$ is the hydrogen position of the minimum of
${\rm V}^{rigid}({\bf r})$ at site "a".
The small polaron Hamiltonian, ${\cal H}_p$, is then the sum of the rigid
lattice Hamiltonian, ${\cal H}$, a perfect lattice
phonon part and the linear coupling between the two,
$$
{\cal H}_p = {\cal H} +
\sum_l \left( {-\hbar^2 \over 2 m_{Pd} } \nabla_l^2
+ { 1 \over 2 } m_{Pd} \omega_l^2 Q_l^2 \right)
+ \sum_l V_l^a Q_l
\eqno(2)
$$
where $m_{Pd}$ is the palladium mass and $\omega_l$ the $l^{th}$ phonon
frequency.
The H-phonon coupling will
cause the phonon coordinates to be displaced by
$V_l^a/m_{Pd} \omega_l^2$.
The energy gained by the phonon displacements is
$ E_{loc}=\sum (V_l^a)^2/2 m_{Pd} \omega_l^2 . $

In the temperature regime where tunneling is the dominant mechanism for
diffusion (T$<$150-100 K) only the ground state will
be populated due to the large gap between the ground and excited states.
The time scale, $\tau_{ab}$, for tunneling from the ground state of site
"a" ($\phi_a$) to any of the three nearest neighbor states "b" ($\phi_b$)
is given by
$$ \tau_{ab}^{-1} = 3 J_{ab} S_{ab} (T) / \hbar . \eqno(3) $$
and is the product of the
Hamiltonian matrix element between adjacent tight-binding states,
$ J_{ab}=\langle \phi_a \vert {\cal H}_p \vert \phi_b \rangle$,
and a temperature
dependent phonon overlap term, $S_{ab}$.
The factor of three accounts for the number of nearest neighbor sites
and the matrix element $J_{ab}$ is equal to 1/6 the groundstate
bandwidth [38].
Phonon coupling destroys some of the coherence of the bands, an effect
which increases with temperature. At T=0,
$$
S_{ab}(T=0)=e^{-\sum_l {1 \over 2}
\left ( {m_{Pd} \omega_l \over 2 \hbar} \right )
\left ( (V_l^a-V_l^b) /m_{Pd} \omega_l \right )^2 }. $$
Just considering the 24 phonon modes from the vibrations of the four atoms
involved in the unit cell plus the four nearest neighbor surface atoms
gives a $E_{loc}$ of 46 meV.
The localization energy is more than half the exact localization
energy (77 meV) gained by relaxing the entire lattice, so the 24 phonon
modes included here contain most of the important relaxation modes.
The phonon displacements that give rise to this localization are small.
For the most strongly coupled mode (a symmetric stretch parallel to the
surface of the three nearest Pd atoms), the phonon displacement,
$V_l^a/m_{Pd} \omega_l^2$, is only 0.08 {\AA}.

The time scale for tunneling from
Equation (3) is 2.6x10$^2$ ps for H, 3.8x10$^4$ ps for D, and 7.2x10$^5$ ps
for T.
The phonon overlap, $S_{ab}$(T=0), is 0.46 and is
only small modification on the tunneling matrix element.
These time scales are all much longer than the sub-picosecond
time scale needed to explain the Hsu, {\it et al.}, experiments within
a quantum delocalization mechanism[8].
The calculated rates are the result of a model potential and are by no
means exact, but there are a few reasons to believe that the tunneling
time scales should be greater than 1 ps.
Since both the diffusion barrier and the lattice constant for
the Ni(111) surface are less than for Pd(111),
the H/Pd(111) bands should be narrower than 4 meV, the calculated
H/Ni(111) groundstate bandwidth [4].
This bandwidth,
when inserted into Equation (3), gives a time scale of .3
ps (without phonon corrections);
a narrower bandwidth would give localization on the helium scattering
time scale.
In addition,
since the isotope dependence is so strong, the experimental results, if
due to large bandwidths, should be different for H and D.
The evidence for delocalization from the Hsu experiments is
the symmetry change with
decreasing coverage from 6-fold to 3-fold and the accompanying
attenuation of the specular intensity.
The delocalized states are deduced using a model for the
helium scattering process in which the helium atoms scatter off
a hydrogen density which in the plane of the surface is a constant value
within an ellipsoid centered at the bridge site and a major semi-axis
equal to 0.8${\rm \AA}$ and minor semi-axis equal to 0.5${\rm \AA}$.
Thus the
hydrogen density has a constant value which spans both sides of the
unit cell and the hydrogen is delocalized within the unit cell.
This is a qualitatively different picture of the hydrogen wavefunction than
the ground state shown in Fig. 1.

There are factors other than the accuracy of the EAM potential which are
relevant in a comparison of the present calculations with the experiments
of Hsu, {\it et al.} The experiments were done at a coverage of a half
so hydrogen-hydrogen interactions might be an issue,
although most evidence supports the conclusion that low
coverage presents the best conditions for delocalization (see above).
Another issue is the presence of subsurface
sites, which are of comparable energies with the surface sites and are
thermally accessible.
These sites, which are not visible to helium scattering, might influence
the symmetry of the surface phase[19].

In summary, the vibrational states for H,D, and T on the Pd(111) surface
were reported. A notable feature of these states is their anharmonicity,
which is most prominent in the weak isotopic dependence of the
parallel vibrational state (see Table I). The poorness of
harmonic estimates of the energies also demonstrates the importance of
anharmonic effects. Harmonic estimates of the energy of the E state of H
are off by 90 meV.  Harmonic estimates do better for the A$_1^1$ state
and get better with increasing mass (see Fig. 2).
The states have a large zero point energy, so it is not surprising that
anharmonic regions of the potential energy surface are important.
The ground state energy for H and the excited state energies for all
three isotopes (see Table I) are greater than the classical barrier between
sites, 175 meV.  That these states can have a high energy relative to a
barrier in one coordinate and still remain largely localized on one side
of this barrier is due to the multidimensional nature of these states.
For a discussion of the localization and zero point/barrier energy
ratio in the context of clusters see Ref. [37].
The bandwidths of the states are narrow and
an analysis of the coupling of the hydrogen eigenstates
to the phonons
indicates that the hydrogen only induces small lattice distortions
(with an amplitude $<$ 0.1 ${\rm \AA}$)
and that the
hydrogen is localized on a particular binding site for a time scale
on the order of a hundred picoseconds or more.
The dynamics of hydrogen as it diffuses across the surface and also
into the subsurface will be the subject of future study[40].

That anharmonic effects are more
important for the computed perpendicular vibration (transitions to the E state)
than the parallel vibration (transitions to the A$_1^1$ state) could
lead to some atypical HREELS results.  HREELS finds that the
the parallel frequency, $\omega_{\parallel}$, is less than the
perpendicular, $\omega_{\perp}$, for the H isotope[17].
Since the perpendicular mode has a stronger isotopic dependence than the
parallel, it is possible that the ordering of the frequencies could
change for heavier hydrogen isotopes and $\omega_{\parallel}$ could
become higher than $\omega_{\perp}$, although the present calculations find
$\omega_{\perp} < \omega_{\parallel}$ for all isotopes.
An HREELS study of H and D on Pd(111) could resolve some of the issues
discussed in this letter,
including the ordering of the parallel and perpendicular
vibrations and the isotopic dependence of these frequencies.
Furthermore, the HREELS spectra could determine whether the
excited states have bandwidth which is much larger
than HREELS resolution (about 1 meV) or of the same magnitude
as we are predicting here.

\vskip 25pt
{\bf Acknowledgements}. The authors are pleased to acknowledge the
assistance of Dr. Diane Lynch, Dr. Lawrence Pratt,
and Professor K. Birgitta Whaley.
We would also like to thank NSF for its support of the present work
through grant CHE 9203498.

\eject
{\bf References}
\vskip 10 pt
\noindent
\item{1. } See, for example, K. Christmann, Surf. Sci. Rep. {\bf 9} (1988) 1.
\item{2. } In 1958, Eigen and De Maeyer proposed a now out of favor
"protonic semiconductor" model for hydrogen in ice crystals.
M. Eigen and L. De Maeyer, Proc. Roy. Soc. A {\bf 247} (1958) 505.
\item{3. } K. Christmann, R.J. Behm, G.Ertl, M.A. Van Hove, and W.H.
Weinberg, J. Chem. Phys. {\bf 70} (1979) 4168.
\item{4. } M.J. Puska, R.M. Nieminen, B. Chakraborty, S. Holloway, and
J.K. N\o rskov, Phys. Rev. Lett. {\bf 51} (1983) 1081; M.J. Puska and R.M.
Nieminen, Surf. Sci. {\bf 157} (1985) 413.
\item{5. } O. Grizzi, M. Shi, H. Bu, J.W. Rabalais, R.R. Rye, and P.
Nordlander, Phys. Rev. Lett. {\bf 63} (1989) 1408.
\item{6. } C. Astaldi, A. Bianco, S. Modesti, and E. Tosatti, Phys. Rev.
Lett. {\bf 68} (1992) 90.
\item{7. } C.M. Mate and G.A. Somorjai, Phys. Rev. B {\bf 34}
(1986) 7417.
\item{8. } C-H Hsu, B.E. Larson, M. El-Batanouny, C.R. Willis, and K.M.
Martini, Phys. Rev. Lett. {\bf 66} (1991) 3164.
\item{9. } M. El-Batanouny, private communication.
\item{10. } S.M. George, A.M. DeSantolo, and R.B. Hall, Surf. Sci. {\bf
159} (1985) L425; D.A. Mullins, B. Roop, and J.M. White, Chem. Phys.
Lett. {\bf 129} (1986) 511; T-S Lin and R. Gomer, Surf. Sci. {\bf 225}
(1991) 41.
\item{11. } R. DiFoggio and R. Gomer, Phys. Rev. B {\bf 25}
(1982) 3490; S.C. Wang and R. Gomer, J. Chem. Phys. {\bf 83} (1985) 4193; C.
Dharmadhikari and R. Gomer, Surf. Sci. {\bf 143} (1984) 223; E.A.
Daniels, J.C. Lin, and R. Gomer, Surf. Sci. {\bf 204} (1988) 129.
\item{12. } E.G. Seebauer, A.C.F. Kong, and L.D. Schmidt, J. Chem. Phys.
{\bf 88} (1988) 6597.
\item{13. } E.G. Seebauer and L.D. Schmidt, Chem. Phys. Lett. {\bf 123}
(1986) 129.
\item{14. } X.D. Zhu, A. Lee, A. Wong, and U. Linke, Phys. Rev. Lett.
{\bf 68} (1992) 1862.
\item{15. } For theoretical examinations of the H/W(110) diffusion
experiments, see K. Freed, J. Chem. Phys. {\bf 82} (1985) 5264;
K.B. Whaley, A. Nitzan, and R.B. Gerber, J. Chem. Phys. {\bf 84}
(1986) 5181;
A. Auerbach, K.F. Freed, and R. Gomer, J.Chem. Phys. {\bf 86} (1987) 2356;
Q. Niu, J. Stat. Phys. {\bf 65} (1991) 317;
P.D. Reilly, R.A. Harris, and K.B. Whaley, J. Chem. Phys. {\bf 97} (1992) 6875.
\item{16. } P-A Karlsson, A-S M\aa rtinson, S Andersson, and P.
Nordlander, Surf. Sci. {\bf 175} (1986) L759.
\item{17. } H. Conrad, M. E. Kordesch, R. Scala, and W. Stenzel,
J. Electron. Spec. and Relat. Phen. {\bf 38} (1986) 289 and
H. Conrad, M. E. Kordesch, W. Stenzel,
and M. Sunjic, J. Vac. Soc. Tech. A {\bf 5} (1987) 452.
\item{18. } M.S. Daw and M.I. Baskes, Phys. Rev. B {\bf 29}
(1984) 6443.
\item{19. } T.E. Felter, S.M. Foiles, M.S. Daw, and R.H. Stulen, Surf.
Sci. {\bf 171} (1986) L379.
\item{20. } H. Conrad, G. Ertl, and E.E. Latta, Surf. Sci. {\bf 4}
(1974) 435.
\item{21. } J. V\"olkl, G. Wollenweber, K.-H. Klatt, and G. Alefeld,
Z. Naturforsch. {\bf 26a} (1971) 922; see also J. V\"olkl and G.
Alefeld, in {\it Diffusion in Solids: Recent Developments}, edited by
A.S. Nowick and J.J. Burton (Academic Press, New York, 1975).
\item{22. } N.W. Ashcroft and N.D. Mermin, {\it Solid State Physics}
(Saunders College, Philadelphia, 1976).
\item{23. } G.E. Kimball and G.H. Shortley, Phys. Rev. {\bf 45}
(1934) 815 and M.J. Puska and R.M. Nieminen, Phys. Rev. B {\bf 29}
(1984) 5382.
\item{24. } D.M. Ceperley and B.J. Alder, Science, {\bf 231}
(1986) 555.
\item{25. } This mass dependence is seen in the tunneling probabilities
for a number of systems, including rectangular and parabolic barriers.
See R.P. Bell, {\it The Tunnel Effect in Chemistry} (Chapman and Hall,
London, 1980).
\item{26. } H. Ibach and D.L. Mills, {\it Electron Energy Loss
Spectroscopy and Surface Vibrations} (Academic Press, New York, 1982).
\item{27. } H. Conrad, R. Scala, W. Stenzel, and R. Unwin, J. Chem.
Phys. {\bf 81} (1984) 6371.
\item{28. }  A.D. Johnson, K.J. Maynard, S.P. Daley, Q.Y. Yang, and S.T.
Ceyer, Phys. Rev. Lett. {\bf 67} (1991) 927; K.J. Maynard, A.D. Johnson,
S.P. Daley, and S.T. Ceyer, Faraday Discuss. Chem. Soc. {\bf 91} (1991) 437.
\item{29. } A.M. Bar\'o, H. Ibach, and H.D. Bruchmann, Surf. Sci. {\bf 88}
(1979) 384.
\item{30. } H. Conrad, W. Stenzel and M.E. Kordesch, presented at the 5th
International Conference on Vibrations at Surfaces, Grainau,
FRG, 1987, p. 0-3.
\item{31. } P.J. Feibelman and D.R. Hamann, Surf. Sci. {\bf 182} (1987) 411.
\item{32. } S. Andersson, Chem. Phys. Lett. {\bf 55} (1978) 185; P.A. Karlson,
A.S. M\aa rtinsson, S. Andersson and P. Nordlander, Surf. Sci. {\bf 175} (1986)
L759.
\item{33. } L.J. Richter,T.A. Germer, J.P. Sethna, and W. Ho,
Phys. Rev. B {\bf 38} (1988) 10403.
\item{34. } C. Nyberg and C.G. Tengst\aa l, Solid State Commun. {\bf 44} (1982)
251; Phys. Rev. Lett. {\bf 50} (1983) 1680.
\item{35. } B. Voigtl\"ander, S. Lehwald, and H. Ibach, Surf. Sci. {\bf 208}
(1989) 113.
\item{36. } D.R. Hamann and P.J. Feibelman, Phys. Rev. B {\bf37} (1988) 3847.
\item{37. } T. Holstein, Ann. Phys. (N.Y.) {\bf 8} (1959) 325; {\bf 8}
(1959) 343; C.P. Flynn and A.M. Stoneham, Phys. Rev. B {\bf 1}
(1970) 3966.
\item{38. } The bandwidth is six times (not twice) the Hamiltonian matirix
element, $\langle \phi_a \vert H \vert \phi_b \rangle$,
because of the three-fold symmetry of the lattice.
\item{39. } S.W. Rick, D.L. Lynch, and J.D. Doll, J. Chem Phys. {\bf
95} (1991) 3506.
\item{40. } S.W. Rick, D.L. Lynch, and J.D. Doll, J. Chem Phys.
(in press).

\vfill\eject
\noindent
Table I. Band centers and band widths for the ground and two first
excited vibrational states of the hydrogen isotopes on the Pd(111)
surface, in meV, and the vibrational frequencies corresponding to the
the A$_1^0$ to E transition ($\omega_\parallel$) and
the A$_1^0$ to A$_1^1$ transition ($\omega_\perp$), in cm$^{-1}$.
$$\vbox{
\settabs 7 \columns
\hrule
\vskip .05 true in
\hrule
\vskip .1 true in
\+ &H& &D& &T& \cr
\vskip .1 true in
\hrule
\vskip .1 true in
\+ & Center & Width & Center & Width&
Center & Width \cr
\+ & (meV) & (meV) & (meV) & (meV) & (meV) & (meV) \cr
\vskip .1 true in
\hrule
\vskip .1 true in
\+ ${\rm A}_1^0$ & 236.3 & 0.011 &169.2 &0.000076 &138.6&
0.000004  \cr
\+ ${\rm A}_1^1$ & 345.9 & 2.17  &252.5 &0.023 &
208.2& 0.0006\cr
\+ ${\rm E}$ & 356.7 & 7.7 &270.7 &0.208 &228.35 &
0.0075 \cr
\vskip .1 true in
\hrule
\vskip .1 true in
\+ Frequencies & & & & & & \cr
\vskip .1 true in
\hrule
\vskip .1 true in
\+ $\omega_\parallel /{\rm cm}^{-1} $ & 971 & &  819 & & 724 \cr
\+ $\omega_\perp /{\rm cm}^{-1} $ & 884 & & 672 & & 561 \cr
\vskip .05 true in
\hrule
\vskip .05 true in
\hrule
}$$
\beginsection {Figure Captions}

\noindent
Figure 1. Contours of the amplitudes of the {\bf k}=0 bandstates
for H/Pd(111) as a function of a
surface coordinate and a coordinate perpendicular to the surface (left
side) and as a function of the two surface coordinates (right side),
for the E (top), A$_1^1$ (middle), and A$_1^0$
states (bottom).
Each rectangle spans an area of 2.75 ${\rm \AA}$ by 4.76 ${\rm \AA}$ and the
coordinate in the [111] direction begins at .02 ${\rm \AA}$ below the surface
plane.

\vskip 15pt
\noindent
Figure 2. Bandstate energies for H, D, and T, comparing the
tight-binding state energies (the longer horizontal lines) to the
harmonic estimates (the shorter horizontal lines).
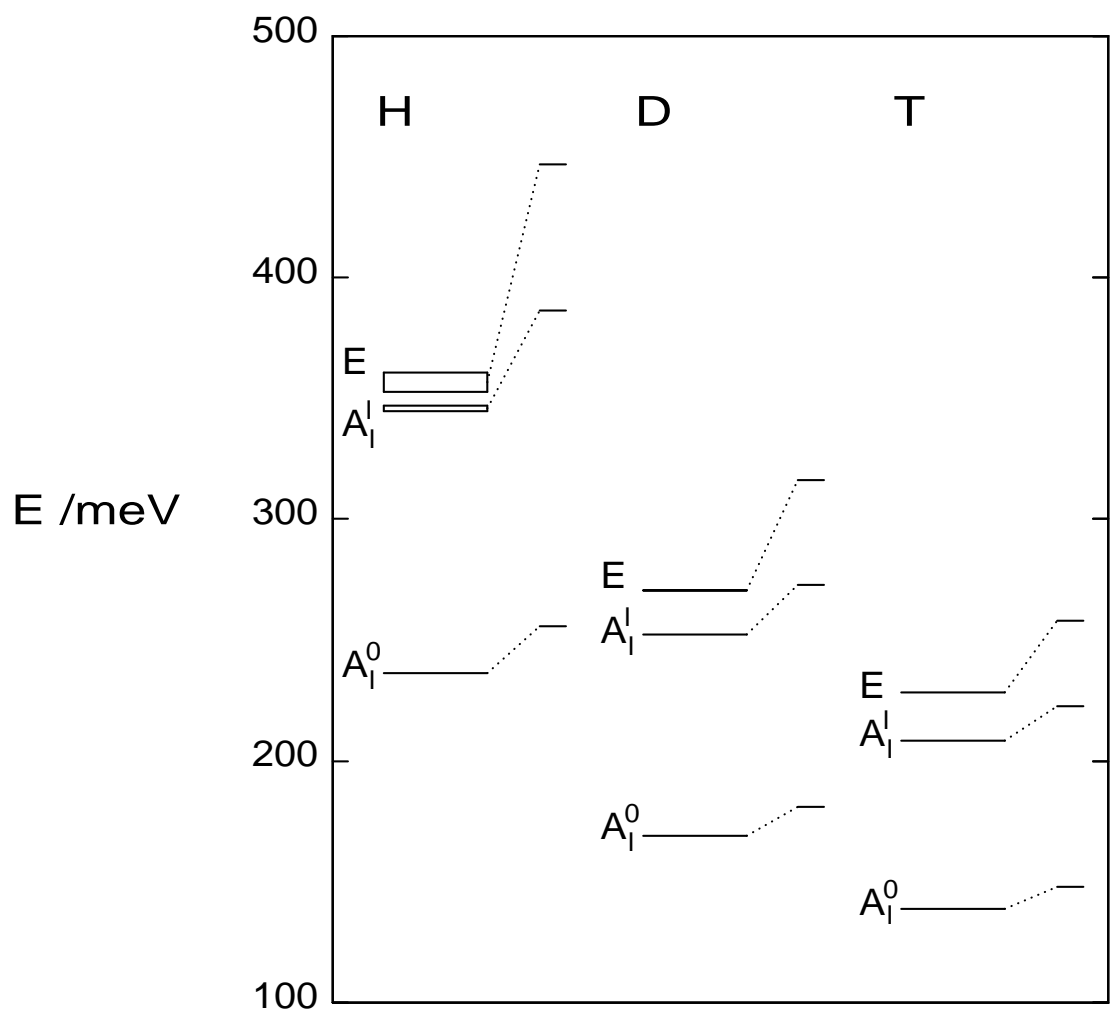
\end